\documentclass[a4paper,10pt,twoside]{cpc-hepnp}
\usepackage{upgreek,fancyhdr}
\usepackage{multicol}
\usepackage{graphicx}
\usepackage{booktabs}
\usepackage{amssymb,bm,mathrsfs,bbm,amscd}
\usepackage[tbtags]{amsmath}
\usepackage{lastpage}
\usepackage{color}
\usepackage{graphicx} 
\usepackage{epstopdf}
\usepackage{hyperref}
\usepackage{subfigure}
\hypersetup{
    colorlinks=true,
    linkcolor=blue,
    filecolor=blue,      
    urlcolor=blue,
    citecolor=blue,
}

\begin{document}
\pdfoutput=1

\title{Three-dimensional potential energy surface for fission of $^{236}$U within covariant density functional theory }

\author{%
 Ming-Hui Zhou$^{1}$, Ze-Yu Li$^{2, 1}$, Sheng-Yuan Chen$^{1}$, \\ 
 Yong-Jing Chen$^{2,\dagger}$ \email{ahchenyj@126.com} and Zhi-Pan Li$^{1,\dagger}$ \email{zpliphy@swu.edu.cn}
}

\maketitle

\address{
$^{1}$ School of Physical Science and Technology, Southwest University, Chongqing 400715, China\\
$^{2}$ China Nuclear Data Center, China Institute of Atomic Energy, Beijing 102413, China\\
}

\begin{abstract}
We have calculated the three-dimensional potential energy surface (PES) for the fission of compound nucleus $^{236}$U using the covariant density functional theory with constraints on the axial quadrupole and octupole deformations $(\beta_2, \beta_3)$ as well as the nucleon number in the neck $q_N$. By considering the additonal degree of freedom $q_N$, coexistence of the elongated and compact fission modes is predicted for $0.9\lesssim \beta_3 \lesssim 1.3$. Remarkably, the PES becomes very shallow across a large range of quadrupole and octupole deformations for small $q_N$, and consequently, the scission line in $(\beta_2, \beta_3)$ plane will extend to a shallow band, which leads to a fluctuation for the estimated total kinetic energies by several to ten MeV and for the fragment masses by several to about ten nucleons.
   
\end{abstract}


\begin{multicols}{2}


\section{\label{sec:level1} INTRODUCTION}

Nuclear fission presents a unique example of non-equilibrium large-amplitude collective motion in a multi-dimensional space where all nucleons participate with complex correlation effects. Fission is considered as one of the most complex processes in nuclear physics and meanwhile offers a rich laboratory for a broad variety of scientific research on nuclear properties and general physics. Therefore, the study of fission always stays at the frontier of nuclear physics. 

The evolution of a nucleus from a single ground-state shape into two separated fragments in nuclear fission has been described in terms of potential energy surfaces (PESs) that are functions of suitable shape coordinates  \cite{Meitner1939,Bohr1939} since its discovery \cite{Hahn1938berDE}. In all the current approaches to fission that rely on the adiabatic approximation, the first step is, therefore, to define the most relevant collective coordinates and compute the PESs. Two types of models, i.e. macroscopic-microscopic (MM) approach \cite{BRACK1972RMP,Nix1972ARNS} and energy density functional theory (DFT) \cite{Schunck2016RPP}, are widely used to calculate the PESs for nuclear fission. 

The MM approach consists in viewing the nucleus as a finite chunk of nuclear matter, the energy of which is parametrized as a function of the charge, mass, and deformations of the nucleus. The total energy includes a macro energy, a shell correction energy, and a pairing energy. The MM approach has a series of versions characterized by different parametrizations of nuclear surface of the liquid drop and different phenomenological nuclear potentials, such as the five-dimensional finite-range liquid-drop model (FRLDM) \cite{Moller2001Nature,Moller2009PRC,Ichikawa2012PRC}, macroscopic-microscopic Woods-Saxon model \cite{Jachimowicz2012PRC,Jachimowicz2013PRC}, the macroscopic–microscopic Lublin–Strasbourg drop (LSD) model in the three-quadratic-surface parametrization \cite{Wang2019CTP,Zhu2020CTP}, the LSD in Fourier shape parametrization \cite{Schmitt2017PRC}, two-center shell model \cite{Liu2019PRC} and so on. Based on the obtained multi-dimensional PESs, various methods for dynamical evolution in the collective space have also been performed and generally reproduced the experimental measurements \cite{Moller2001Nature,Randrup2011PRL,Aritomo2014PRC,Mumpower2020PRC,Pomorski2021CPC,Liu2021PRC,Verriere2021PRC,Sierk2017PRC}.

Self-consistent approaches based on the nuclear DFT have recently demonstrated that a microscopic description has a potential for both qualitative and quantitative description of fission data \cite{Schunck2016RPP,Schmidt2018RPP,Simenel2018PPNP,Bender2020JPG,Verriere2020FP,Schunck2022PPNP}. In the DFT framework, the wave functions along the fission path are generally determined by the minimization of the energy of the nucleus within a given set of constraints and assumed symmetries. Nowdays, large-scale calculations for fission up to scission are generally performed in two dimensions, e.g. the axially symmetric quadrupole and octupole deformations, and have been done based on the non-relativistic Gogny and Skyrme energy density functionals (EDFs) \cite{Berger1984NPA,Warda2002PRC,Goutte2005PRC,Dubray2008PRC,Staszczak2009PRC,Younes2009PRC,Sadhukhan2016PRC,Schunck2014PRC,Regnier2016PRC,Regnier2019PRC,Verriere2021PRC2,Chen2022CPC} and also the relativistic (covariant) EDFs \cite{TaoH2017PRC,LiZY2022PRC,ZhaoJ2022PRC,Ren2022PRC,ZhaoJ2019PRC,ZhaoJ2019PRC2}. For instance, Dubray $et~al.$ \cite{Dubray2008PRC} have used the constrained Hartree-Fock-Bogoliubov (HFB) method with the Gogny D1S functional to calculate PESs of fissioning nuclei $^{226}$Th and $^{256,258,260}$Fm as functions of the quadrupole and octupole moments. Corresponding fragment properties at scission are derived yielding fragment deformations, deformation energies, energy partitioning, neutron multiplicities, total fragment kinetic energies and so on. Schunck $et~al.$ \cite{Schunck2014PRC} employed the Skyrme EDFs: SkM$^*$, UNEDF0, and UNEDF1 to calculate the PESs of $^{240}$Pu in different two-dimensional spaces. In particular, the effects of the triaxial degree of freedom and the nucleon number in the neck on the scission have been intensively studied. We have also performed a constrained relativistic mean-field calculations in the collective space of axially symmetric quadrupole and octupole deformations, based on the energy density functional PC-PK1, to determine the PES, scission line, collective masses etc. for the induced fission of $^{226}$Th \cite{TaoH2017PRC}.

The self-consistent approach guarantees that the shape for a given set of constraints is optimal, although nonconstrained degrees of freedom are not fixed. However, the results provided with this method are not always unique \cite{Dubray2012CPC} as one can easily land for a given set of constraints in one of the local minima. As a consequence, discontinuities are often observed in the PESs in the restricted deformation space, especially for the large elongated configurations. In fact, these discontinuities are entirely spurious since locally enlarging the collective space can easily restore the continuity of the full PES \cite{Dubray2012CPC}. In addition, continuous PESs give additional flexibility to define the scission configurations and improve the predictive power of the theory. Very recently, microscopic calculations of PESs in a fully three-dimensional (3D) collective space for fission up to scission have been performed, e.g. PES of $^{228}$Th in axial quadrupole and octupole deformations as well as an isoscalar pairing degree of freedom \cite{ZhaoJ2021PRC}, and PESs of $^{252}$Cf and $^{258}$No in axial quadrupole, octupole, and hexadecapole moments \cite{Zdeb2021PRC}. 

In this work, we will take the compound nucleus $^{236}$U as an example to calculate the fully 3D PES in axial quadrupole and octupole deformations $(\beta_2, \beta_3)$ as well as the nucleon number in the neck $q_N$ based on the covariant DFT (CDFT). It has been well demonstrated that the $q_N$ degree of freedom provides a mechanism to pass continuously from a single whole nucleus to two distinct fragments \cite{Younes2009PRC,Schunck2014PRC,Han2021PRC}. Therefore, here we will mainly focus on the impact of the additional degree of freedom $q_N$ on the fission modes and scission configurations. Sec.~\ref{sec:level2} introduces the theoretical framework briefly. The details of the calculation and the results
for potential energy surfaces, scission configurations, and estimated total kinetic energies are described and discussed in Sec.~\ref{sec:level3}. Sec.~\ref{sec:level4} contains a summary of results and an outlook for future studies.
\


\section{\label{sec:level2} CONSTRAINED COVARIANT DENSITY FUNCTIONAL THEORY}

The energy density functional in the point-coupling version for the CDFT can be written as
\begin{gather}
  \begin{split}
  {{E}_{\rm CDF}}&=\int{{d}}\mathbf{r}{{\varepsilon }_{\rm CDF}}(\mathbf{r}) \\ 
             &=\sum\limits_{k}{\int{d}}\mathbf{r}\upsilon _{k}^{2}{{{\bar{\psi }}}_{k}}(\mathbf{r})(-i\mathbf{\gamma }\mathbf{\nabla} +m){{\psi }_{k}}(\mathbf{r}) \\ 
             &+\int{d\mathbf{r}\left( \frac{{{\alpha }_{S}}}{2}\rho _{S}^{2}+\frac{{{\beta }_{S}}}{3}{{\rho}_S^{3}}+\frac{{{\gamma }_{S}}}{4}\rho _{S}^{4}+\frac{{{\delta }_{S}}}{2}{{\rho }_{S}}\Delta {{\rho }_{S}} \right.} \\ 
             &+\frac{{{\alpha }_{V}}}{2}{{j}_{\mu }}{{j}^{\mu }}+\frac{{{\gamma }_{V}}}{4}{{({{j}_{\mu }}{{j}^{\mu }})}^{2}}+\frac{\delta_V}{2}{{j}_{\mu }}\Delta {{j}^{\mu }}+\frac{{e}}{2}{{\rho }_{p}}{{A}^{0}} \\ 
             &\left. +\frac{{{\alpha }_{TV}}}{2}j_{TV}^{\mu }\cdot {{(j_{TV})}_{\mu }}+\frac{{{\delta }_{TV}}}{2}j_{TV}^{\mu }\cdot \Delta (j_{TV})_{\mu } \right) \label{EDF}
  \end{split}  
\end{gather}
with the local densities and currents
\begin{gather}
  \begin{split}
  {{\rho }_{S}}(\mathbf{r})&=\sum\limits_{k}{v_{k}^{2}}{{\bar{\psi }}_{k}}(\mathbf{r}){{\psi }_{k}}(\mathbf{r})\\
  {{j}^{\mu }}(\mathbf{r})&=\sum\limits_{k}{v_{k}^{2}}{{\bar{\psi }}_{k}}(\mathbf{r}){{\gamma }^{\mu }}{{\psi }_{k}}(\mathbf{r})\\
  j_{TV}^{\mu }(\mathbf{r})&=\sum\limits_{k}{v_{k}^{2}}{{\bar{\psi }}_{k}}(\mathbf{r}){{\gamma }^{\mu }}\tau_3{{\psi }_{k}}(\mathbf{r})
\end{split}
\end{gather}
where $\psi$ is the Dirac spinor of the nucleon. $\rho_p$ and $A^0$ are respectively the proton density and Coulomb field. Coupling constants $(\alpha, \beta, \gamma, \delta)$ are determined by PC-PK1 parametrization\cite{Zhao2010PRC} in this work. The subscripts indicate the symmetry of the couplings: $S$ stands for scalar, $V$ for vector, and $T$ for isovector.

By means of the variation of the energy density functional with respect to the densities and currents, one can then obtain the relativistic Kohn–Sham equation, which has the form of a single-particle Dirac equation,
\begin{gather}
  \{-i\mathbf{\alpha}\cdot\mathbf{\nabla} +V(\mathbf{r})+\beta [M+S(\mathbf{r})]\}\psi_k(\mathbf{r})=\varepsilon_i\psi_k(\mathbf{r}) \label{Dirac equ1}
\end{gather}
The single-particle effective Hamiltonian contains local scalar $S(\mathbf{r})$ and vector $V(\mathbf{r})$ potentials, which are functions of densities and currents
\begin{gather}
  \begin{split}
  S(\mathbf{r})&=\alpha_S \rho_S+\beta_S\rho_S^2+\gamma_S\rho_S^3+\delta_S\Delta\rho_S\\
  V^\mu (\mathbf{r})&=\alpha_V j^\mu+\gamma_V(j_\nu j^\nu)j^\mu+\delta_V\Delta j^\mu+eA^\mu\frac{1-\tau_3}{2}\\
           & \ \ \ +\tau_3(\alpha_{TV}j^\mu_{TV}+\delta_{TV}\Delta j_{TV}^\mu)
  \end{split}
\end{gather}

Pairing correlations between nucleons are treated using the Bardeen-Cooper-Schrieffer (BCS) approach with a $\delta$ pairing force \cite{Burvenich2002PRC}. Due to the broken of the translational symmetry, one has to consider the center-of-mass (c.m.) correction energy for the motion of the c.m. and here a phenomenological formulas $E_{\rm c.m.}=-\frac{3}{4}\cdot 41A^{-1/3}$ is adopted. Finally, the total energy reads
\begin{gather}
  E_{\rm tot}=E_{\rm CDF}+E_{\rm pair}+E_{\rm c.m.}
\end{gather}

To calculate the multi-dimensional PES in a large deformation space, one needs to solve the Dirac equation (\ref{Dirac equ1}) with high precision and efficiency. One way is to expand the Dirac spinor in a two-center harmonic oscillator (TCHO) basis which are eigen functions in a TCHO potential
\begin{gather}
  V\left(r_{\perp}, z\right)=\frac{1}{2} M \omega_{\perp}^{2} r_{\perp}^{2}+ \begin{cases}\frac{1}{2} M \omega_{1}^{2}\left(z+z_{1}\right)^{2}, & z<0 \\ \frac{1}{2} M \omega_{2}^{2}\left(z-z_{2}\right)^{2}, & z \geq 0\end{cases}
\end{gather}
in $z$ direction of the cylindrical coordinate system. TCHO can be regarded as two off-center harmonic oscillators connected at $z=0$, while $z_1 (z_2)$ and $\omega_1 (\omega_2)$ denote the distance from $z=0$ to the center of the left (right) harmonic oscillator and its frequency, respectively. Details can be found in Refs. \cite{Geng2007CPL,LiZY2023}.

The entire map of the energy surface in three-dimensional collective space for fission is obtained by imposing constraints on the three collective coordinates: quadrupole deformation $\beta_2$, octupole deformation $\beta_3$, and the number of nucleons in the neck $q_N$
\begin{equation}
  \langle E_{\rm tot}\rangle+\sum\limits_{k=2,3}C_k(\langle\hat Q_k\rangle-q_k)^2+C_N(\langle\hat Q_N\rangle-q_N)^2,
\end{equation}
where $\langle E_{\rm tot}\rangle$ is the total energy of CDFT, $\hat Q_2$, $\hat Q_3$, and $\hat Q_N$ denote the mass quadrupole and octupole operators, and the Gaussian neck operator, respectively. $q_k$ and $q_N$ are the constraint values of these operators. The Gaussian neck operator is generally chosen as $\hat Q_N=\exp[-(z-z_N)^2/a^2_N]$, where $a_N$ = 1 fm and $z_N$ is the position of the neck determined by minimizing $\langle\hat Q_N\rangle$ \cite{Younes2009PRC}. The left and right fragments are defined as parts of the whole nucleus with $z\leq z_N$ and $z\geq z_N$, respectively. 

Once the constraint on $q_N$ is adopted, the variation of the configurations around scission will become smooth and continuous. Therefore, the Coulomb energy between left and right fragments calculated by 
\begin{equation}
  \label{Eq:EC}
 E_{\rm C} (\beta_2, \beta_3, q_N)=e^2\int d\mathbf{r} d\mathbf{r^\prime} \frac{\rho^L_p(\mathbf{r})\rho^R_p(\mathbf{r^\prime})}{|\mathbf{r}-\mathbf{r^\prime}|},
\end{equation}
where $\rho_p^L$ ($\rho_p^R$) is the proton density of the left (right) fragment at the configuration $(\beta_2, \beta_3, q_N)$, will be also smooth around scission and can be used to estimate the total kinetic energy (TKE).



\section{\label{sec:level3}RESULTS AND DISCUSSION}

In this section we present the results of an illustrative study of 3D PES for the fission of compound nucleus $^{236}$U. In the first step a large-scale deformation-constrained CDFT calculation is performed to generate the two-dimensional (2D) PES in the $(\beta_2, \beta_3)$ plane. The range of collective variables is -1.02$-$7.34 for $\beta_{2}$ with a step $\Delta \beta_{2} = 0.08$, and from 0.00$-$3.76 for $\beta_{3}$ with a step $\Delta \beta_{3}$ = 0.08. Then we extend the 2D PES to 3D by the constraint on $q_N$ from a thick neck to as thinner as possible with a step $\Delta q_N = 0.5$ for the configurations with neck in the 2D calculation. The energy density functional PC-PK1 \cite{Zhao2010PRC} is used for the effective interaction in the particle-hole channel, and a $\delta$-force pairing with strengths: $V_n = 344$ MeV fm$^3$ and $V_p = 371$ MeV fm$^3$ determined by the empirical pairing gap parameters of $^{236}$U, calculated using a five-point formula \cite{Bender2000EPJA}. The self-consistent Dirac equation for the single-particle wave functions is solved by expanding the nucleon spinors in an axially deformed TCHO basis in cylindrical coordinates with 20 major shells.

\begin{center}
    \includegraphics[scale=0.4]{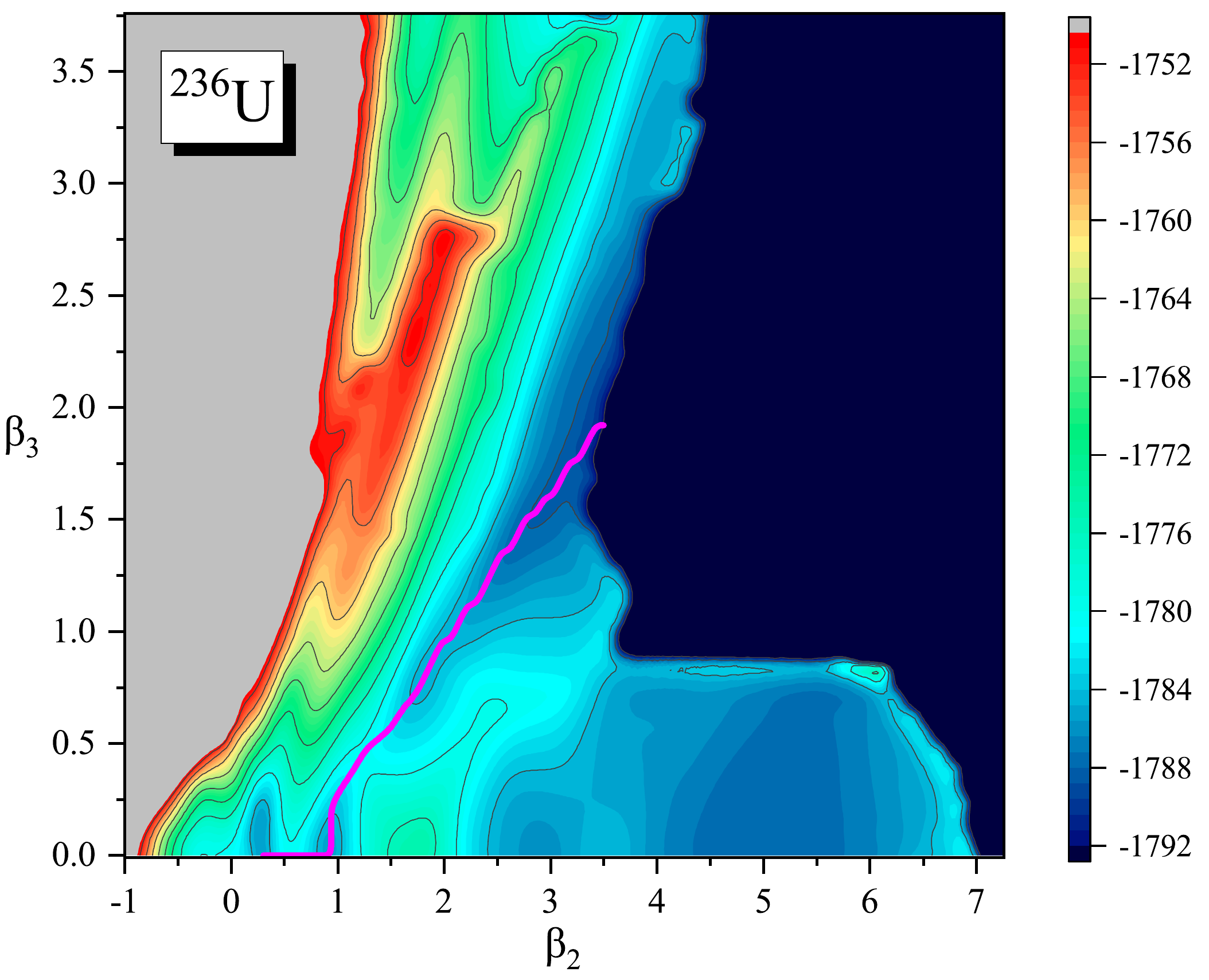}
    \figcaption{\label{fig:Vcoll} (color online) Potential energy surface of $^{236}$U in the
    $(\beta_{2}, \beta_{3})$ plane calculated by the constrained CDFT with PC-PK1 functional. The magenta solid line denotes the optimal fission path in $(\beta_2, \beta_3)$ plane. The energy difference between neighboring contour lines is 4.0 MeV.} 
 \end{center}

\begin{figure*}[htbp]
  \centering
 \includegraphics[scale=0.77]{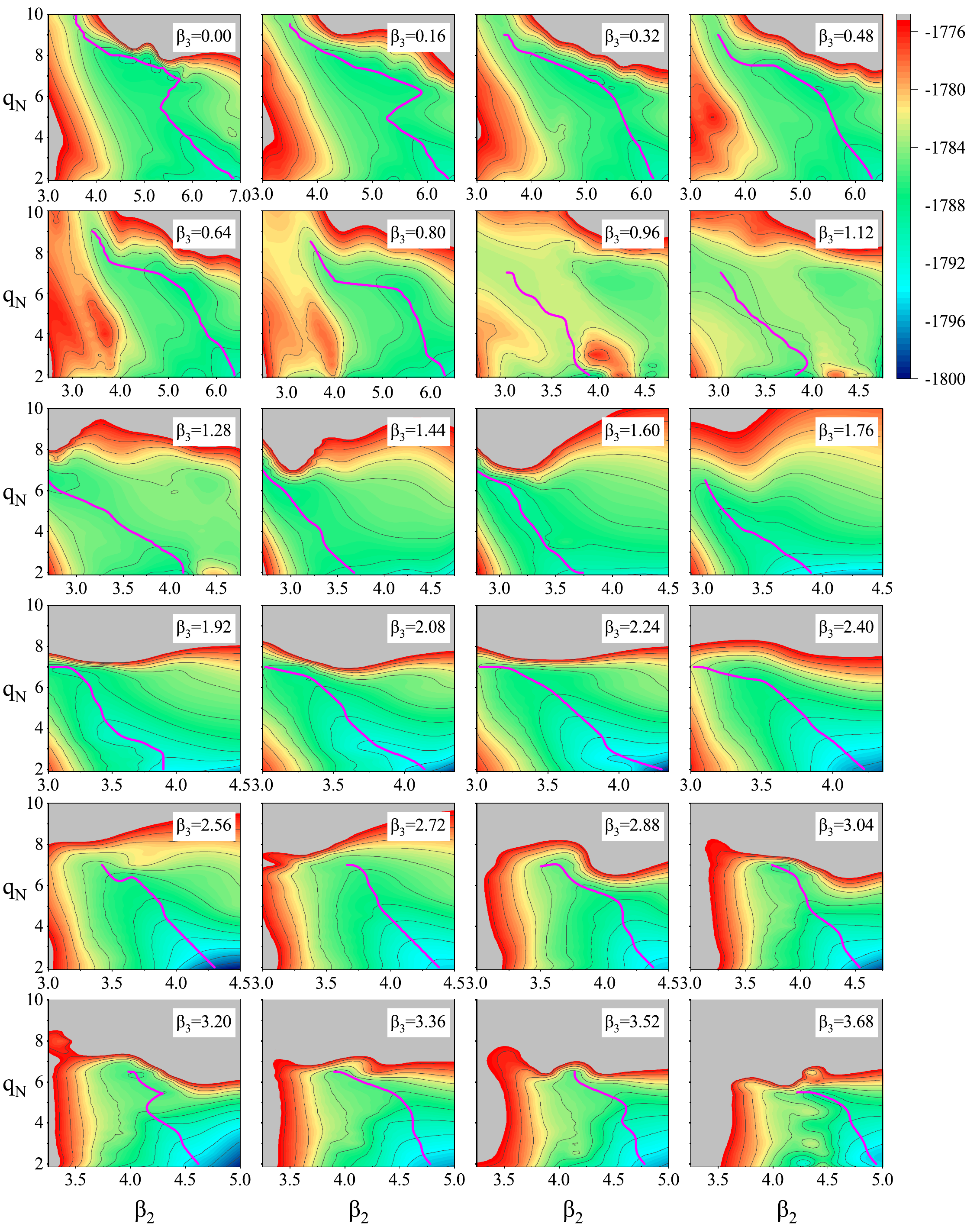}
  \caption{\label{fig:3D} (Color online) Contour plots for the sections of the 3D PES of $^{236}$U. In each panel the energy is shown as a function of $\beta_2$ and $q_{N}$ when $\beta_{3}$ is fixed at a certain value. The magenta solid lines are the optimal fission paths in $(\beta_2, q_N)$ planes. The energy difference between neighboring contour lines is 2.0 MeV.}
\end{figure*}

Figure \ref{fig:Vcoll} displays the 2D PES calculated by constrained CDFT with PC-PK1 functional. The equilibrium shape is located at $(\beta_{2}, \beta_{3})\sim (0.30, 0.00)$. A double-humped fission barrier is predicted along the static fission path, and the calculated heights are 6.85 and 6.05 MeV for the inner and outer fission barriers, respectively. At $\beta_2 > 2.0$  a symmetric valley up to very large elongation is found, and the symmetric and asymmetric fission valleys are separated by a ridge from $(\beta_2, \beta_3) \approx  (1.7, 0.0)$ to $(3.5, 1.0)$. In addition, one can also see a fission valley heading towards large octupole deformation starting from the equilibrium shape. It  describes superasymmetric fission strongly related to cluster radioactivity and the saddle point in the superasymmetric valley reaches more than $30$ MeV. It is noted that the overall topography of the 2D PES is similar to that calculated with  the Skyrme SkM* functional \cite{Schunck2021Book}.

\begin{center}
  \includegraphics[scale=0.65]{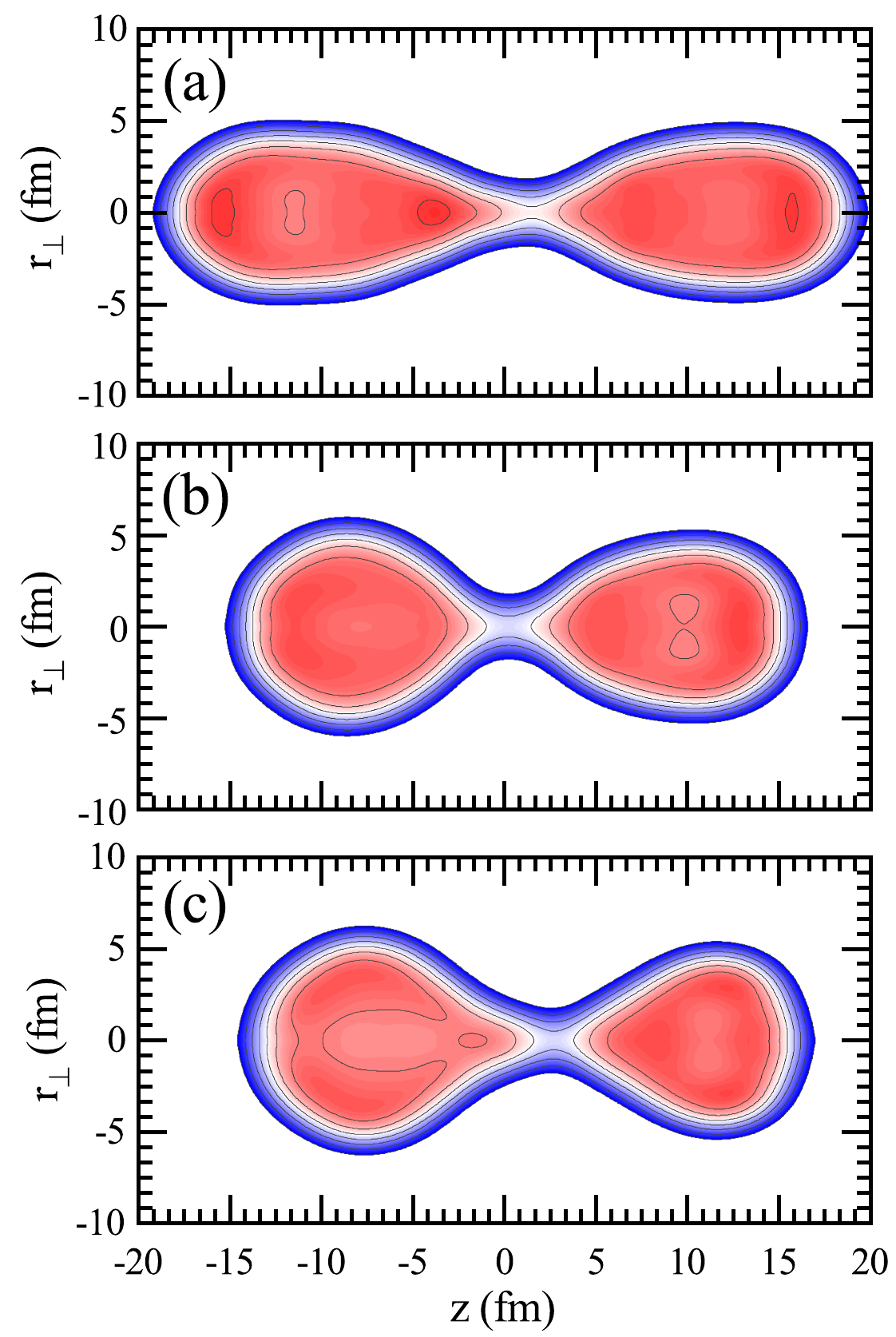}
  \figcaption{\label{fig:density} (color online) The density distributions of $^{236}$U for the configurations with $q_N=2$ but $(\beta_2, \beta_3) = (6.22, 0.80)$ (a), $(3.90, 0.96)$ (b), and $(3.90, 2.08)$ (c).} 
\end{center}

To visualize a 3D PES is a difficult task that can somehow be facilitated if one of the variables is kept fixed and the PES for the remaining two variables is plotted as a contour plot. Such a procedure has been followed in Fig. \ref{fig:3D} for $^{236}$U, where we show sections of the PES for fixed values of $\beta_3$ as maps in the $(\beta_2, q_N)$ space. Here we mainly focus on the region from saddle to scission. The magenta solid lines are the optimal fission paths in $(\beta_2, q_N)$ planes. As a whole, the PES descends towards large elongation and thin neck as shown by the magenta solid line. Remarkably, the PES broadens up to form a wide ``estuary'' in the $(\beta_2, q_N)$ subspace for $q_N<6$: the energy surface is very shallow across a large range of quadrupole deformations. In fact, we have also checked that this wide ``estuary'' is also existent in the $(\beta_3, q_N)$ subspace for small $q_N$. This should manifest itself by a sizable broadening of the yields and total kinetic energies, which will be discussed in detail in Fig. \ref{fig:scission}. 

Specifically, for $\beta_3\lesssim  0.80$, the 2D PESs in $(\beta_2, q_N)$ plane are all extended to very large quadrupole deformations $\beta_2>6.0$ and the density distribution for one of the configurations is shown in Fig. \ref{fig:density} (a). Obviously, both the fissioning nucleus and its pre-fragments are very elongated. As the asymmetric octupole deformation $\beta_3$ increases to 0.96, two fission valleys corresponding to the compact fission mode and elongated fission mode coexist and they are separated by a shallow ridge. The optimal fission mode is the compact one, whose density distribution is also illustrated in Fig. \ref{fig:density} (b). It is well known that the compact mode is driven by the double shell closure $Z=50$ and $N=82$. The coexistence of these two fission valleys lasts to $\beta_3=1.28$, which just corresponds to the end of the ridge that separates the symmetric and asymmetric fission valleys in the 2D PES in Fig. \ref{fig:Vcoll}. When $\beta_3>1.44$, the pattern of contour plots is similar but only the optimal fission paths shift to larger $\beta_2$. There are mainly two fission modes in this region: the compact one shown in Fig. \ref{fig:density} (b) for $\beta_3\lesssim 2.0$ and the one shown in Fig. \ref{fig:density} (c) characterized by octupole-deformed pre-fragments \cite{Scamps2018Nature} for the region with larger $\beta_3$.

\begin{center}
  \includegraphics[scale=0.75]{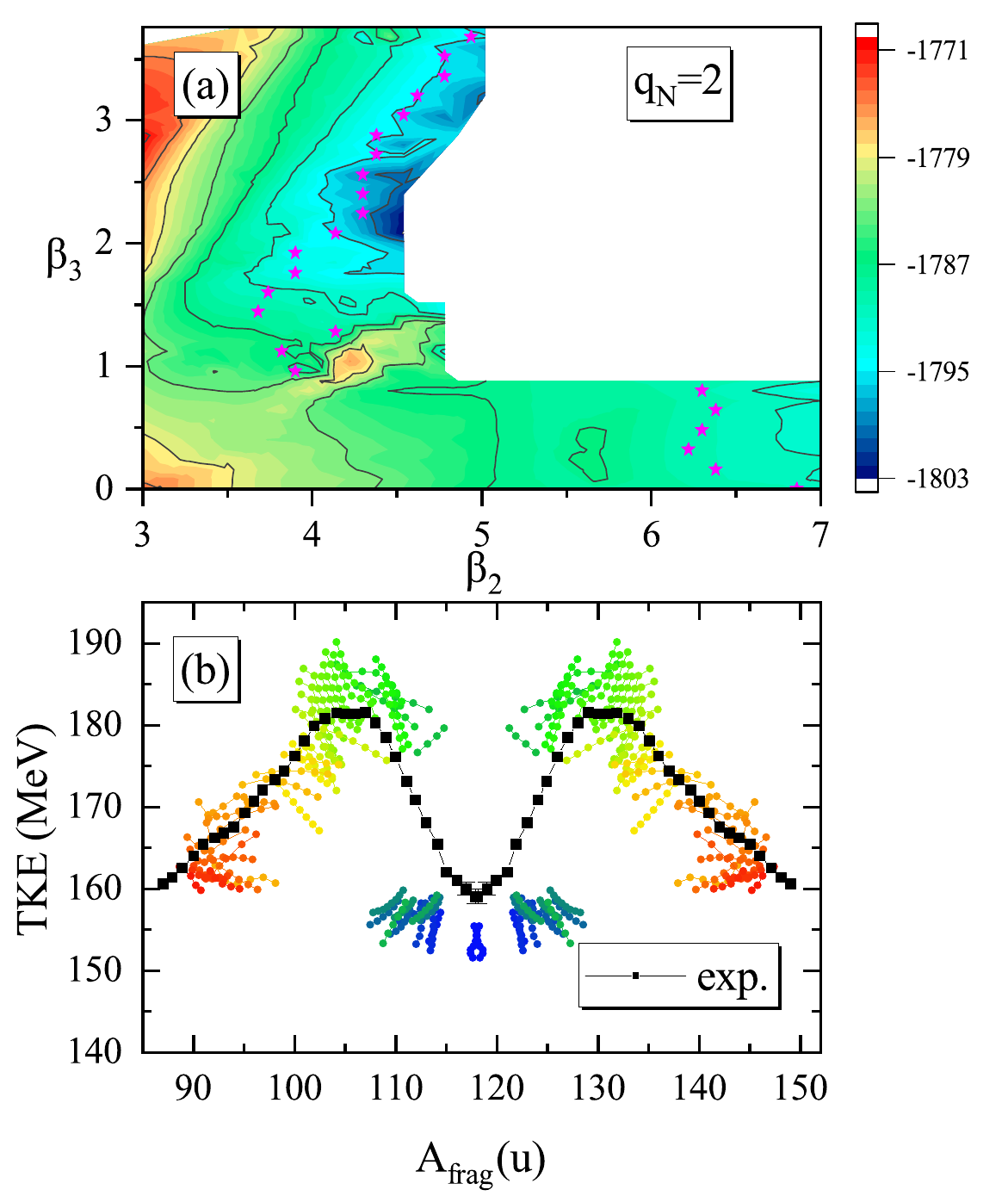}
  \figcaption{\label{fig:scission} (color online) (a) Contour plot for the section of 3D PES with fixed $q_N=2$. The star symbols denote the configurations on the optimal fission paths as shown in Fig. \ref{fig:3D}. The energy difference between neighboring contour lines is 4.0 MeV. (b) The calculated Coulomb energies between two fragments using Eq. (\ref{Eq:EC}) to estimate the TKEs for the configurations with energies lower than $E_\star+1$ MeV, where $E_\star$ is the total binding erergy of the star symbol in panel (a). The experimental TKE distribution for the thermal neutron induced fission of $^{235}$U has also been shown for comparison.} 
\end{center}

To domenstrate the impact of the additional degree of freedom $q_N$ on the scissions and the consequent fragment masses and TKEs, we have shown the  contour plot for the section of 3D PES with fixed $q_N=2$ and the calculated Coulomb energies between two fragments using Eq. (\ref{Eq:EC}) to estimate the TKEs for the configurations around the optimal scission points in Fig. \ref{fig:scission}. In panel (a), we find that the energy surface is very shallow around the optimal scission points (star symbols), especially to the larger elongations, for both the symmetric and asymmetric fission valleys. Within $\pm 1$ MeV, the quadrupole and octupole deformations can extend even to $\sim 0.5$ unit. This will for sure broaden the distributions of the yields and TKEs, which is quantitatively estimated in Fig. \ref{fig:scission} (b). For a certain fragment partitioning, the estimated TKE can fluctuate about several to ten MeV or even more than ten MeV at the mass of heavy fragment $A_{\rm H}\sim 132$. Moreover, for $122\lesssim A_{\rm H}\lesssim  128$, one notes that the lower-than 160 MeV and higher-than 178 MeV are simultaneously obtained since the coexistence of the elongated and compact fission modes in the region with $0.9\lesssim \beta_3\lesssim  1.3$ (c.f. Fig. \ref{fig:3D}). The measured TKE could be an average of those of different fission modes with corresponding probabilities. Finally, we would like to emphasize that, for the asymmetric fission with $A_{\rm H}>130$, the broaden of the energy surface around scission can lead to a fluctuation of the fragment masses by several to about ten nucleons, which is consistent with the width of the asymmetric peak of the yield distribution in the actinides.



\section{\label{sec:level4}SUMMARY}

In summary, we have performed a fully three-dimensional calculation to generate the PES for the fission of compound nucleus $^{236}$U using the constrained covariant density functional theory with constraints on the axial quadrupole and octupole deformations $(\beta_2, \beta_3)$ as well as the nucleon number in the neck $q_N$. By considering the additonal degree of freedom $q_N$, coexistence of the elongated and compact fission modes is predicted for $0.9\lesssim \beta_3 \lesssim 1.3$, and remarkably, the PES broadens up to form a wide ``estuary'' in the $(\beta_2, q_N)$ subspace for $q_N<6$: the energy surface is very shallow across a large range of quadrupole deformations. This wide ``estuary'' is also existent in the $(\beta_3, q_N)$ subspace for small $q_N$. Therefore, the scission line in $(\beta_2, \beta_3)$ plane will extend to a shallow band, and consequently, leads to a fluctuation for the estimated TKE by several to ten MeV and for the fragment masses by several to about ten nucleons. Of course, this is just a simple estimation for the fluctuation of the fission observables. More precise calculation should be done by performing dynamical evolution of the collective wave packet on the 3D PES, e.g. using the time-dependent generator coordinate method \cite{Regnier2018CPC}. Such work is in progress.



\vspace{0.5cm}

\bibliography{cpcbib}

\providecommand{\noopsort}[1]{}\providecommand{\singleletter}[1]{#1}%
\begin{thebibliography}{10}

\bibitem{Meitner1939}
L.~Meitner and Otto~Robert Frisch.
\newblock Disintegration of uranium by neutrons: a new type of nuclear
  reaction.
\newblock {\em Nature}, 143:239--240, 1939.

\bibitem{Bohr1939}
Niels Bohr and John~Archibald Wheeler.
\newblock The mechanism of nuclear fission.
\newblock {\em Phys. Rev.}, 56:426--450, Sep 1939.

\bibitem{Hahn1938berDE}
O.~Th. Hahn and Fritz Strassmann.
\newblock {\"U}ber die entstehung von radiumisotopen aus uran durch bestrahlen
  mit schnellen und verlangsamten neutronen.
\newblock {\em Naturwissenschaften}, 26:755--756, 1938.

\bibitem{BRACK1972RMP}
M.~BRACK, JENS DAMGAARD, A.~S. JENSEN, H.~C. PAULI, V.~M. STRUTINSKY, and C.~Y.
  WONG.
\newblock Funny hills: The shell-correction approach to nuclear shell effects
  and its applications to the fission process.
\newblock {\em Rev. Mod. Phys.}, 44:320--405, Apr 1972.

\bibitem{Nix1972ARNS}
J~R Nix.
\newblock Calculation of fission barriers for heavy and superheavy nuclei.
\newblock {\em Annual Review of Nuclear Science}, 22(1):65--120, 1972.

\bibitem{Schunck2016RPP}
N.~Schunck and L.~M. Robledo.
\newblock Microscopic theory of nuclear fission: A review.
\newblock {\em Reports on Progress in Physics}, 79(11), 10 2016.

\bibitem{Moller2001Nature}
P~Moller, DG~Madland, AJ~Sierk, and A~Iwamoto.
\newblock Nuclear fission modes and fragment mass asymmetries in a
  five-dimensional deformation space.
\newblock {\em NATURE}, 409(6822):785--790, FEB 15 2001.

\bibitem{Moller2009PRC}
Peter M\"oller, Arnold~J. Sierk, Takatoshi Ichikawa, Akira Iwamoto, Ragnar
  Bengtsson, Henrik Uhrenholt, and Sven \AA{}berg.
\newblock Heavy-element fission barriers.
\newblock {\em Phys. Rev. C}, 79:064304, Jun 2009.

\bibitem{Ichikawa2012PRC}
Takatoshi Ichikawa, Akira Iwamoto, Peter M\"oller, and Arnold~J. Sierk.
\newblock Contrasting fission potential-energy structure of actinides and
  mercury isotopes.
\newblock {\em Phys. Rev. C}, 86:024610, Aug 2012.

\bibitem{Jachimowicz2012PRC}
P.~Jachimowicz, M.~Kowal, and J.~Skalski.
\newblock Secondary fission barriers in even-even actinide nuclei.
\newblock {\em Phys. Rev. C}, 85:034305, Mar 2012.

\bibitem{Jachimowicz2013PRC}
P.~Jachimowicz, M.~Kowal, and J.~Skalski.
\newblock Eight-dimensional calculations of the third barrier in ${}^{232}$th.
\newblock {\em Phys. Rev. C}, 87:044308, Apr 2013.

\bibitem{Wang2019CTP}
Zhi-Ming Wang, Wen-Jie Zhu, Xin Zhu, Chun-Lai Zhong, and Tie-Shuan Fan.
\newblock 236u multi-modal fission paths on a five-dimensional deformation
  surface.
\newblock {\em Communications in Theoretical Physics}, 71(4):417, apr 2019.

\bibitem{Zhu2020CTP}
Xin Zhu, Zhi-Ming Wang, Wen-Jie Zhu, Chun-Lai Zhong, Yi-Mo Zhang, Long-Yong
  Liao, and Tie-Shuan Fan.
\newblock Macroscopic–microscopic calculations of fission potential surface
  of uranium isotopes in the three quadratic surfaces parametrization.
\newblock {\em Communications in Theoretical Physics}, 72(10):105301, sep 2020.

\bibitem{Schmitt2017PRC}
C.~Schmitt, K.~Pomorski, B.~Nerlo-Pomorska, and J.~Bartel.
\newblock Performance of the fourier shape parametrization for the fission
  process.
\newblock {\em Phys. Rev. C}, 95:034612, Mar 2017.

\bibitem{Liu2019PRC}
Li-Le Liu, Xi-Zhen Wu, Yong-Jing Chen, Cai-Wan Shen, Zhu-Xia Li, and Zhi-Gang
  Ge.
\newblock Study of fission dynamics with a three-dimensional langevin approach.
\newblock {\em Phys. Rev. C}, 99:044614, Apr 2019.

\bibitem{Randrup2011PRL}
J\o{}rgen Randrup and Peter M\"oller.
\newblock Brownian shape motion on five-dimensional potential-energy surfaces:
  Nuclear fission-fragment mass distributions.
\newblock {\em Phys. Rev. Lett.}, 106:132503, Mar 2011.

\bibitem{Aritomo2014PRC}
Y.~Aritomo, S.~Chiba, and F.~Ivanyuk.
\newblock Fission dynamics at low excitation energy.
\newblock {\em Phys. Rev. C}, 90:054609, Nov 2014.

\bibitem{Mumpower2020PRC}
M.~R. Mumpower, P.~Jaffke, M.~Verriere, and J.~Randrup.
\newblock Primary fission fragment mass yields across the chart of nuclides.
\newblock {\em Phys. Rev. C}, 101:054607, May 2020.

\bibitem{Pomorski2021CPC}
Krzysztof Pomorski, José~M. Blanco, Pavel~V. Kostryukov, Artur Dobrowolski,
  Bożena Nerlo-Pomorska, Michał Warda, Zhi-Gang Xiao, Yong-Jing Chen, Li-Le
  Liu, Jun-Long Tian, Xin-Yue Diao, and Qiang-Hua Wu.
\newblock Fission fragment mass yields of th to rf even-even nuclei *.
\newblock {\em Chinese Physics C}, 45(5):054109, may 2021.

\bibitem{Liu2021PRC}
Li-Le Liu, Yong-Jing Chen, Xi-Zhen Wu, Zhu-Xia Li, Zhi-Gang Ge, and Krzysztof
  Pomorski.
\newblock Analysis of nuclear fission properties with the langevin approach in
  fourier shape parametrization.
\newblock {\em Phys. Rev. C}, 103:044601, Apr 2021.

\bibitem{Verriere2021PRC}
Marc Verriere and Matthew~Ryan Mumpower.
\newblock Improvements to the macroscopic-microscopic approach of nuclear
  fission.
\newblock {\em Phys. Rev. C}, 103:034617, Mar 2021.

\bibitem{Sierk2017PRC}
Arnold~J. Sierk.
\newblock Langevin model of low-energy fission.
\newblock {\em Phys. Rev. C}, 96:034603, Sep 2017.

\bibitem{Schmidt2018RPP}
Karl-Heinz Schmidt and Beatriz Jurado.
\newblock {Review on the progress in nuclear fission\textemdash{}experimental
  methods and theoretical descriptions}.
\newblock {\em Rept. Prog. Phys.}, 81(10):106301, 2018.

\bibitem{Simenel2018PPNP}
C.~Simenel and A.~S. Umar.
\newblock {Heavy-ion collisions and fission dynamics with the time-dependent
  Hartree\textendash{}Fock theory and its extensions}.
\newblock {\em Prog. Part. Nucl. Phys.}, 103:19--66, 2018.

\bibitem{Bender2020JPG}
Michael Bender et~al.
\newblock {Future of Nuclear Fission Theory}.
\newblock {\em J. Phys. G}, 47(11):113002, 2020.

\bibitem{Verriere2020FP}
Marc Verriere and David Regnier.
\newblock The time-dependent generator coordinate method in nuclear physics.
\newblock {\em Frontiers in Physics}, 8, 2020.

\bibitem{Schunck2022PPNP}
Nicolas Schunck and David Regnier.
\newblock Theory of nuclear fission.
\newblock {\em Progress in Particle and Nuclear Physics}, 125:103963, 2022.

\bibitem{Berger1984NPA}
JF~BERGER, M~GIROD, and D~GOGNY.
\newblock Microscopic analysis of collective dynamics in low-energy fission.
\newblock {\em NUCLEAR PHYSICS A}, 428(OCT):C25--C36, 1984.

\bibitem{Warda2002PRC}
M.~Warda, J.~L. Egido, L.~M. Robledo, and K.~Pomorski.
\newblock Self-consistent calculations of fission barriers in the fm region.
\newblock {\em Phys. Rev. C}, 66:014310, Jul 2002.

\bibitem{Goutte2005PRC}
H.~Goutte, J.~F. Berger, P.~Casoli, and D.~Gogny.
\newblock Microscopic approach of fission dynamics applied to fragment kinetic
  energy and mass distributions in $^{238}\mathrm{U}$.
\newblock {\em Phys. Rev. C}, 71:024316, Feb 2005.

\bibitem{Dubray2008PRC}
N.~Dubray, H.~Goutte, and J.-P. Delaroche.
\newblock Structure properties of $^{226}\mathrm{Th}$ and
  $^{256,258,260}\mathrm{Fm}$ fission fragments: Mean-field analysis with the
  gogny force.
\newblock {\em Phys. Rev. C}, 77:014310, Jan 2008.

\bibitem{Staszczak2009PRC}
A.~Staszczak, A.~Baran, J.~Dobaczewski, and W.~Nazarewicz.
\newblock Microscopic description of complex nuclear decay: Multimodal fission.
\newblock {\em Phys. Rev. C}, 80:014309, Jul 2009.

\bibitem{Younes2009PRC}
W.~Younes and D.~Gogny.
\newblock Microscopic calculation of $^{240}\mathrm{Pu}$ scission with a
  finite-range effective force.
\newblock {\em Phys. Rev. C}, 80:054313, Nov 2009.

\bibitem{Sadhukhan2016PRC}
Jhilam Sadhukhan, Witold Nazarewicz, and Nicolas Schunck.
\newblock Microscopic modeling of mass and charge distributions in the
  spontaneous fission of pu-240.
\newblock {\em PHYSICAL REVIEW C}, 93(1), JAN 20 2016.

\bibitem{Schunck2014PRC}
N.~Schunck, D.~Duke, H.~Carr, and A.~Knoll.
\newblock Description of induced nuclear fission with skyrme energy
  functionals: Static potential energy surfaces and fission fragment
  properties.
\newblock {\em Phys. Rev. C}, 90:054305, Nov 2014.

\bibitem{Regnier2016PRC}
D.~Regnier, N.~Dubray, N.~Schunck, and M.~Verri\`ere.
\newblock Fission fragment charge and mass distributions in
  $^{239}\mathrm{Pu}(n,f)$ in the adiabatic nuclear energy density functional
  theory.
\newblock {\em Phys. Rev. C}, 93:054611, May 2016.

\bibitem{Regnier2019PRC}
D.~Regnier, N.~Dubray, and N.~Schunck.
\newblock {From asymmetric to symmetric fission in the fermium isotopes within
  the time-dependent generator-coordinate-method formalism}.
\newblock {\em Phys. Rev. C}, 99(2):024611, 2019.

\bibitem{Verriere2021PRC2}
Marc Verriere, Nicolas Schunck, and David Regnier.
\newblock Microscopic calculation of fission product yields with
  particle-number projection.
\newblock {\em PHYSICAL REVIEW C}, 103(5), MAY 3 2021.

\bibitem{Chen2022CPC}
Yong-Jing Chen, Yang Su, Guoxiang Dong, Li-Le Liu, Zhigang Ge, and Xiaobao
  Wang.
\newblock Energy density functional analysis of the fission properties of
  240pu: The effect of pairing correlations.
\newblock {\em Chinese Physics C}, 46(2):024103, feb 2022.

\bibitem{TaoH2017PRC}
H.~Tao, J.~Zhao, Z.~P. Li, T.~Nik\ifmmode \check{s}\else
  \v{s}\fi{}i\ifmmode~\acute{c}\else \'{c}\fi{}, and D.~Vretenar.
\newblock Microscopic study of induced fission dynamics of $^{226}\mathrm{Th}$
  with covariant energy density functionals.
\newblock {\em Phys. Rev. C}, 96:024319, Aug 2017.

\bibitem{LiZY2022PRC}
Zeyu Li, Shengyuan Chen, Yongjing Chen, and Zhipan Li.
\newblock Microscopic study on asymmetric fission dynamics of
  $^{180}\mathrm{Hg}$ within covariant density functional theory.
\newblock {\em Phys. Rev. C}, 106:024307, Aug 2022.

\bibitem{ZhaoJ2022PRC}
Jie Zhao, Tamara Nik\ifmmode \check{s}\else \v{s}\fi{}i\ifmmode~\acute{c}\else
  \'{c}\fi{}, and Dario Vretenar.
\newblock Time-dependent generator coordinate method study of fission. ii.
  total kinetic energy distribution.
\newblock {\em Phys. Rev. C}, 106:054609, Nov 2022.

\bibitem{Ren2022PRC}
Z.~X. Ren, J.~Zhao, D.~Vretenar, T.~Nik\ifmmode \check{s}\else
  \v{s}\fi{}i\ifmmode~\acute{c}\else \'{c}\fi{}, P.~W. Zhao, and J.~Meng.
\newblock Microscopic analysis of induced nuclear fission dynamics.
\newblock {\em Phys. Rev. C}, 105:044313, Apr 2022.

\bibitem{ZhaoJ2019PRC}
Jie Zhao, Tamara Nik\ifmmode \check{s}\else \v{s}\fi{}i\ifmmode~\acute{c}\else
  \'{c}\fi{}, Dario Vretenar, and Shan-Gui Zhou.
\newblock Microscopic self-consistent description of induced fission dynamics:
  Finite-temperature effects.
\newblock {\em Phys. Rev. C}, 99:014618, Jan 2019.

\bibitem{ZhaoJ2019PRC2}
Jie Zhao, Jian Xiang, Zhi-Pan Li, Tamara Nik\ifmmode \check{s}\else
  \v{s}\fi{}i\ifmmode~\acute{c}\else \'{c}\fi{}, Dario Vretenar, and Shan-Gui
  Zhou.
\newblock Time-dependent generator-coordinate-method study of mass-asymmetric
  fission of actinides.
\newblock {\em Phys. Rev. C}, 99:054613, May 2019.

\bibitem{Dubray2012CPC}
N.~Dubray and D.~Regnier.
\newblock Numerical search of discontinuities in self-consistent potential
  energy surfaces.
\newblock {\em Computer Physics Communications}, 183(10):2035--2041, 2012.

\bibitem{ZhaoJ2021PRC}
Jie Zhao, Tamara Nik\ifmmode \check{s}\else \v{s}\fi{}i\ifmmode~\acute{c}\else
  \'{c}\fi{}, and Dario Vretenar.
\newblock Microscopic self-consistent description of induced fission: Dynamical
  pairing degree of freedom.
\newblock {\em Phys. Rev. C}, 104:044612, Oct 2021.

\bibitem{Zdeb2021PRC}
A.~Zdeb, M.~Warda, and L.~M. Robledo.
\newblock Description of the multidimensional potential-energy surface in
  fission of $^{252}\mathrm{Cf}$ and $^{258}\mathrm{No}$.
\newblock {\em Phys. Rev. C}, 104:014610, Jul 2021.

\bibitem{Han2021PRC}
R.~Han, M.~Warda, A.~Zdeb, and L.~M. Robledo.
\newblock Scission configuration in self-consistent calculations with neck
  constraints.
\newblock {\em Phys. Rev. C}, 104:064602, Dec 2021.

\bibitem{Zhao2010PRC}
P.~W. Zhao, Z.~P. Li, J.~M. Yao, and J.~Meng.
\newblock New parametrization for the nuclear covariant energy density
  functional with a point-coupling interaction.
\newblock {\em Phys. Rev. C}, 82:054319, Nov 2010.

\bibitem{Burvenich2002PRC}
T.~B\"urvenich, D.~G. Madland, J.~A. Maruhn, and P.-G. Reinhard.
\newblock Nuclear ground state observables and qcd scaling in a refined
  relativistic point coupling model.
\newblock {\em Phys. Rev. C}, 65:044308, Mar 2002.

\bibitem{Geng2007CPL}
Geng Li-Sheng, Meng Jie, and Toki Hiroshi.
\newblock Reflection asymmetric relativistic mean field approach and its
  application to the octupole deformed nucleus 226ra.
\newblock {\em Chinese Physics Letters}, 24(7):1865, jul 2007.

\bibitem{LiZY2023}
Zeyu Li, Shengyuan Chen, Yongjing Chen, and Zhipan Li.
\newblock Convariant density functional theory in two-center harmonic
  oscillator basis for fission.
\newblock {\em In preparison}, 2023.

\bibitem{Bender2000EPJA}
M~Bender, K~Rutz, PG~Reinhard, and JA~Maruhn.
\newblock Pairing gaps from nuclear mean-field models.
\newblock {\em EUROPEAN PHYSICAL JOURNAL A}, 8(1):59--75, MAY 2000.

\bibitem{Schunck2021Book}
Nicolas Schunck, Zachary Matheson, and David Regnier.
\newblock Microscopic calculation of fission fragment mass distributions at
  increasing excitation energies.
\newblock In Jutta Escher, Yoram Alhassid, Lee~A. Bernstein, David Brown, Carla
  Fr{\"o}hlich, Patrick Talou, and Walid Younes, editors, {\em Compound-Nuclear
  Reactions}, pages 275--284, Cham, 2021. Springer International Publishing.

\bibitem{Scamps2018Nature}
Guillaume Scamps and Cedric Simenel.
\newblock Impact of pear-shaped fission fragments on mass-asymmetric fission in
  actinides.
\newblock {\em NATURE}, 564(7736):382--385, DEC 20 2018.

\bibitem{Regnier2018CPC}
D.~Regnier, N.~Dubray, M.~Verri\`ere, and N.~Schunck.
\newblock Felix-2.0: New version of the finite element solver for the time
  dependent generator coordinate method with the gaussian overlap
  approximation.
\newblock {\em Comput. Phys. Commun.}, 225:180--191, 2018.

\end{thebibliography}
\bibliographystyle{unsrt}

\end{multicols}

\end{document}